# Non-contact *in situ* multi-diagnostic NMR/dielectric spectroscopy

Alysson F. Morais[a,b], Sambhu Radhakrishnan[a,b], Gavriel Arbiv[a,b,c], Dirk Dom[a,b], Karel Duerinckx[a,b], Vinod Chandran C. [a,b], Johan A. Martens[a,b] and Eric Breynaert[a,b,c],*

[a] Centre for Surface Chemistry and Catalysis – Characterization and Application Team (COK-KAT), Celestijnenlaan 200F Box 2461, 3001-Heverlee, Belgium
[b] NMR for Convergence Research (NMRCoRe), KU Leuven, Celestijnenlaan 200F Box 2461, 3001-Heverlee, Belgium.
[c] Center for Molecular Water Science (CMWS), Notkestraße 85, 22607 Hamburg, Germany

*Email: eric.breynaert@kuleuven.be

**ABSTRACT**

Introduction of a dielectric material in an NMR probe head modifies the frequency response of the probe circuit, a phenomenon revealed by the detuning of the probe. For NMR spectroscopy, this detuning is corrected for by tuning and matching the probe head prior to the NMR measurement. The magnitude of the probe detuning - 'the dielectric shift' - provides direct access to the dielectric properties of the sample, enabling NMR spectrometers to simultaneously perform both dielectric and NMR spectroscopy. By measuring sample permittivity as function of frequency, permittivity spectroscopy can be performed using the new methodology. As a proof concept, this was evaluated on methanol, ethanol, 1-propanol, 1-pentanol and 1-octanol using a commercial CPMAS NMR probe head. The results accurately match literature data collected by standard dielectric spectroscopy techniques. Subsequently, the method was also applied to investigate the solvent-surface interactions of water confined in the micropores of an MFI-type, hydrophilic zeolite with Si/Al ratio of 11.5. In the micropores, water adsorbs to Brønsted acid sites and defect sites, resulting in a drastically decreased dielectric permittivity of the nano-confined water. A theoretical background for the new methodology is provided using an effective electric circuit model of a CPMAS probe head with solenoid coil, describing the detuning resulting from insertion of dielectric samples in the probe head.

# INTRODUCTION

Many chemical, biochemical and physical processes occur at solid-solvent interfaces. Characterization of interfacial interactions is however extremely tedious, especially in 3D porous materials such as zeolites, mesoporous silicates, metal organic frameworks (MOFs) and covalent organic frameworks (COFs). Such materials are nevertheless abundantly implemented in chemical applications,[1–3] in molecular water research[4,5] and for energy storage applications[6–12]. Specifically for zeolites, solvent (e.g. water, alcohol) pore wall interactions impact catalysis and adsorption processes.[13,14]

Solvent-pore wall interactions drastically influence the intermolecular organization of the solvent. Near surfaces, water can form a low-density state with a high degree of hydrogen bonding frustration.[15–17] Stabilized by a high electrical potential barrier preventing reorientation of its dipole moment, water molecules tend to align parallel to the surface forming ordered (ice-like) phases.[16,18] Such ordered water exhibits very low polarizability, since the reorientation upon applying an electric field is limited by the surface interaction. This results in permittivity values as low as 2.[18] In hydrophilic nanopores, phase-pure cubic ice $I_c$ is observed at moderate pressure and temperature, owing to the stabilizing influence exerted on the molecular organization by the solid-water interface.[19] Here again, water features low permittivity values due to the interaction with the adsorption sites in the solid.[20]

While physical-chemical phenomena in confined environment clearly are of great technological and scientific importance,[8–12,21,22] investigation of confined systems is challenging as confinement also introduces difficulties for the physicochemical characterization, especially for solvent/pore-wall interactions. In contrast with methods such as neutron and X-ray diffraction,[23–26] for which the reduced length scale of periodicity imposed by nano-confinement is problematic, solid-state nuclear magnetic resonance (NMR) has demonstrated its merit for the investigation of nano-confined systems. Not only is NMR characterization nearly unaffected by the length scale of periodicity, with the advent of high-pressure magic-angle spinning (MAS),[27,28] also high-pressure *in situ* MAS NMR studies of nano-confined water systems are within reach. NMR probes local chemical composition and provides molecular level information on molecular dynamics and intermolecular interactions, including those of and with the confining host material.[14,29–31] It yields absolute quantification ($^1$H, $^2$H, $^{13}$C, $^{29}$Si, $^{17}$O,…) and probes the local chemical environment of $H_2$ and $H_2O$ molecules and their deuterated forms.[32–34] Whereas NMR is focused to local atomic level characterization, dielectric relaxation spectroscopy (DRS), the time-domain variant of electrochemical impedance spectroscopy (EIS), yields medium- to long-range information on the confined water phase and its phase transitions.[4] For water confined in hydrophobic mesoporous (organo-)silicas, for example, DRS revealed a non-freezable water layer at the pore-wall/water interface, demonstrating layering of the confined water.[5] Water molecules in the interfacial layer featured higher activation energies and longer dielectric relaxation times compared to the water pool in the center of the pore, which resembled bulk water.

While NMR and DRS investigations of confined systems are relatively abundant in the literature, correlation of the information acquired with individual techniques is very often difficult. With the here proposed NMR/dielectric multi-diagnostic approach, the long range and average information given by DRS can be seamlessly correlated to the molecular level picture obtained by NMR, uniquely providing the data required to rationalize these effects of host-guest interactions in nanopores. The method allows to record sample permittivity over the range of NMR frequencies both in static and MAS conditions, thus allowing for simultaneous *in situ* NMR and permittivity spectroscopy.

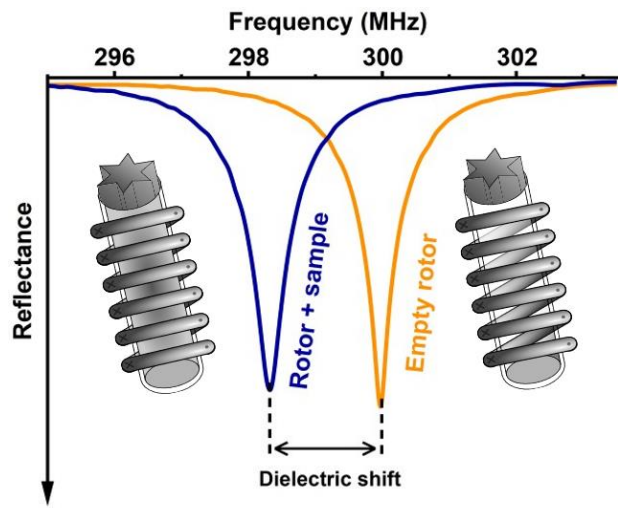

**Figure 1**. Insertion of a sample in an NMR coil causes the detuning of the probe, the dielectric shift. This effect is seen from the reflectance curve of the probe circuit, also known as wobble curve.

Standard MAS NMR probe heads are based on a resonant circuit containing a solenoid coil. For an NMR experiment to work properly, the resonance frequency of the circuit should match the Larmor frequency ($\omega_{NMR} = \gamma B_0$) defined by the static magnetic field ($\boldsymbol{B_0}$) and the gyromagnetic ratio ($\gamma$) of the targeted nucleus. Also, to achieve maximum detection efficiency, the impedance of the probe head circuit should be matched to 50 Ω. When a sample with dielectric properties different from air is inserted in the NMR coil, the frequency response of the probe circuit is modified, as revealed by the detuning of the probe (**Figure 1**). This detuning, hereafter named dielectric shift, is corrected for by tuning and matching the probe prior to NMR measurements and, in general, not much importance is given to its magnitude. Here, it is demonstrated that this detuning can be exploited to gain access to the dielectric properties of the sample, transforming NMR probe heads into multi-diagnostic characterization tools allowing to simultaneously perform permittivity and NMR spectroscopy. As a proof of concept of this method, the permittivity spectra of methanol, ethanol, 1-propanol, 1-pentanol and 1-octanol have been recorded and compared to literature data acquired using standard DRS techniques. The new method was also applied to characterize water confined in an MFI-type zeolite, demonstrating the potential of multi-diagnostic NMR/dielectric spectroscopy for investigating confined solvent-pore wall interactions. Asides confined water systems, the new *in situ* multi-diagnostic will also benefit the fields of battery research,[35–37] food quality control[38,39] and sensing[40], where both NMR and EIS are common characterization techniques.

**EXPERIMENTAL**

**Materials.** MFI-11.5 (CBV2314, MFI framework type, Si/Al = 11.5) was purchased from Zeolyst International in Ammonium-form. Methanol, ethanol 1-propanol, 1-pentanol and 1-octanol were purchased from Fisher Scientific (United Kingdom) and used without further purification or dehydration. Rotor caps with full shaft and o-ring were bought from Rototec (Germany).

**Load impedance measurement**. The load impedance of a H/BB double resonance 4 mm solid-state magic angle spinning (MAS) 300 MHz Bruker probe head was measured in the reflectance mode using a HP 8712C vector network analyser (VNA) connected to the $^1$H channel of the probe via a 50 Ω ($Z_0$) coaxial cable (see also **Discussion S1** and **Figure S1** in the Supporting Information). To enable a direct measurement of the load impedance ($Z_L$) of the probe head, automatically accounting for the phase shift introduced by the leads during the VNA measurements, the electrical delay ($\tau_{ED}$ = 16.818 ns) resulting from the leads was determined and configured in the VNA. To determine the electrical delay, the phase trace of the reflected signal was analysed in the range from 200 to 400 MHz and the electric delay set in the VNA was changed until the phase trace became mostly flat, implying the linear phase delay introduced by the leads was successfully subtracted. With the proper electrical delay set in the VNA, an empty zirconia 4 mm NMR rotor was inserted in the probe and spun at a MAS rate of 2 kHz. The probe was then tuned to 300.131 MHz and matched to 50 Ω. After ejection of the rotor, the load impedance of the probe head was measured in the frequency range from 297.0 to 303.0 MHz, as shown in **Figure S2**. Alternatively to this calibration approach, the load impedance can also be calculated from the input impedance ($Z_{in}$, no electrical delay) as:

$$Z_L(Z_{in}) = Z_0 \frac{Z_{in} - iZ_0 \tan(\pi \tau_{ED} f)}{Z_0 - iZ_{in} \tan(\pi \tau_{ED} f)} \quad (Eq.\,1)$$

**Calibration curves for dielectric permittivity measurements.** To use an NMR probe for dielectric permittivity measurements, for each frequency, the dielectric shift has to be measured for a set of samples with known dielectric constant. To construct these calibration curves, an empty 4 mm zirconia NMR rotor was inserted in the NMR probe head and the probe was tuned to the frequency of interest and matched to 50 Ω. Ultrapure water has been used as permittivity standard, as it features a constant dielectric permittivity $\varepsilon_{H_2O} = 79$ up to 1 GHz at 22 °C.[41] To mimic samples with different dielectric constants, different amounts of Milli-Q water were added to a 4 mm zirconia NMR rotor capped with a vespel cap with a solid shaft and o-ring. The o-ring prevents leakage, while a solid shaft ensures that a well-defined cylindrical water layer is formed inside the rotor under spinning. The total internal volume of such a rotor is $V_R$ = 104 μL. When a volume $V_{H_2O}$ of water is inserted in the rotor, the average permittivity of the content of the rotor is $\varepsilon_{av} = 1 + 79 V_{H_2O}/V_R$. Calibration curves (dielectric shift *vs* dielectric permittivity) were constructed at frequencies from 75 to 800 MHz attained at the 300 MHz, 500 MHz, and 800 MHz Bruker probe heads (**Figure S3**). Using the 300 MHz probe head as an example, the dielectric shift was modelled as a function of the permittivity of the sample, the NMR receiver coil has 8.5 turns, a diameter of 5 mm and is 10 mm long. The dielectric shift is here defined by the detuning caused by the insertion of the sample in the coil taking as a reference the situation in which an empty rotor is inserted in the probe head (see **Figure 1**).

**Dielectric permittivity of alcohols.** Proof of concept measurements of the permittivity spectrum of methanol, ethanol 1-propanol, 1-pentanol and 1-octanol at frequencies from 75 to 800 MHz at 22 °C were recorded using the NMR probehead circuit. *Ca.* 104 μL of each alcohol was added to a 4 mm zirconia NMR rotor capped with an o-ring vespel cap with a solid shaft. No magic-angle spinning was applied. The temperature was regulated at 22 °C with a Bruker BCU unit.

**Dielectric permittivity of water under confinement.** To demonstrate the applicability of the presented methodology to investigate dielectric properties of nanoconfined water, the dielectric and the spectroscopic properties of water adsorbed in an MFI-11.5 zeolite with Si/Al ratio of 11.5 was monitored. The ammonium-form of the zeolite was calcined at 550 °C for 8 h to promote conversion to its H-form. The zeolite was packed into a 4 mm zirconia rotor and dried for 16 h at 200 °C under vacuum (75 mTorr). Following dehydration, the zeolite was loaded with water and the rotor was capped with a vespel cap with a solid shaft and o-ring, avoiding leakage or additional adsorption of water from the air. The capped rotor with the hydrated sample was equilibrated overnight at 60 °C prior to the measurement. To determine the effective dielectric permittivity of the water fraction in the sample, the dielectric shift of the dehydrated zeolite was subtracted from the dielectric shift of the hydrated sample. A 4 mm H/X/Y triple resonance CPMAS solid-state MAS Bruker probe ($^1$H Larmor Frequency 500 MHz) head was employed in these measurements and its calibration curve for dielectric shift (**Figure S4**) was used for the permittivity determination.

**NMR spectroscopy.** $^1$H direct excitation NMR spectra were acquired on a Bruker Avance III 500 MHz (11.7 T) spectrometer equipped with a 4 mm H/X/Y triple resonance CPMAS solid-state MAS probe. Quantitative $^1$H direct excitation spectra were recorded with a π/2 radio-frequency pulse (RF) at 83 kHz, averaging 8 transients with a recycle delay of 2 s.[33] The spectra were referenced to secondary reference adamantane $^1$H resonance at 1.81 ppm, which was further referenced to primary reference, TMS at 0 ppm. Spectral decomposition was performed with DMFIT[42] software.

## RESULTS & DISCUSSION

**NMR Probe Circuit and Calibration.** When an oscillatory electric signal is passed through a solenoidal NMR coil, axial magnetic ($B_1$) and electric fields ($E_1$) are generated inside the coil.[43–45] As the field lines of $E_1$ pass through everything enclosed inside the coil, changes in the dielectric properties of coil contents (e.g., the sample) exert an immediate effect on the circuit. To provide further evidence that the detuning of the probe (the dielectric shift, as illustrated and defined in **Figure 1**) is caused by the electric interaction between sample and receiver coil, the modelling of a probe head circuit as seen from its $^1$H channel was performed. The modelled probe is a H/BB double resonance 4 mm solid-state MAS 300 MHz Bruker probe head in which the $^1$H matching is performed through a variable capacitor in series with the receiver coil and the tuning is performed through a shorted quarter-wave transmission line with variable length that is in parallel with the coil (see **Figure S5**). The length of this quarter-wave transformer can be varied in the vicinity of one quarter of the wavelength of the $^1$H signal. An effective circuit describing the effect of $E_1$ on the resonant properties of the modelled probe head can be drawn as depicted in **Figure 2** (inset), with the self-inductance of the coil being represented by an ideal inductor $L_0 = \mu_0 N^2 \pi r^2 / l$, with $N$ the number of turns, $r$ the radius and $l$ the length of the solenoid. For the modelled probe, $N$ = 8.5, $r$ = 5 mm, $l$ = 10 mm and $L_0$ = 178 nH. More details on how this circuit model is a minimum representation of the real probe circuit can be found in the Supporting Information (see **Discussion S2**).

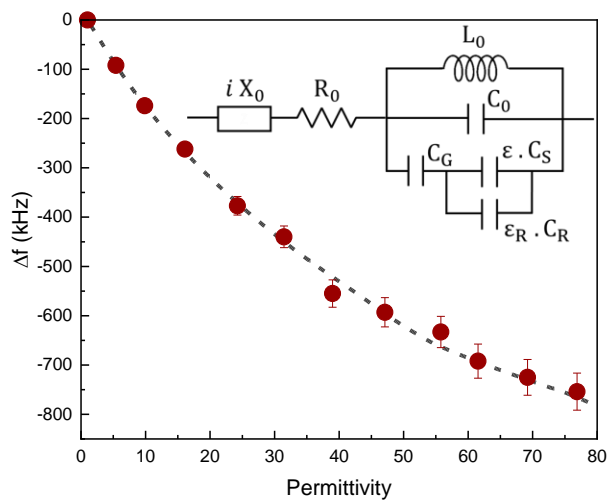

**Figure 2**. Dielectric shift (Δf) as a function of the average permittivity (ε) of the sample inserted in the coil for a 300 MHz Bruker probe head. The curve was constructed by loading the NMR rotor with different amounts of water (see Experimental section). Error bars represent a typical 5% standard deviation obtained from 5 measurements of a rotor filled with ultra-pure water. The dashed lines represent the fit to the equivalent circuit (inset) used to model the resonance frequency of the NMR probe.

Inside the solenoid, the field lines of $E_1$ pass through the air and across portions of the rotor and the sample, contributing to the stray capacitance of the coil. This contribution is taken into account by the capacitors $C_G$, $C_R$ and $C_S$, respectively. As $E_1$ is axial and homogeneous inside the coil, $C_R$ and $C_S$ can be modelled as parallel plate capacitors filled with dielectric materials with dielectric permittivity $\varepsilon_R$ and $\varepsilon$, respectively representing the dielectric constants of zirconia and the sample inside the rotor. $C_R$ and $C_S$ can be written as[44,45]:

$$C_S = \frac{\epsilon_0 \pi d_{RIn}^2}{4\,l}, \qquad (Eq.\,2)$$

$$C_R = \frac{\epsilon_0 \pi (d_{REx}^2 - d_{RIn}^2)}{4\,l} \qquad (Eq.\,3)$$

with $\epsilon_0$ the vacuum permittivity and $d_{RIn}$ = 2.95 mm and $d_{REx}$ = 4.0 mm the internal and external radii of the NMR rotor.

Experimental investigation of resonant properties and stray capacitances in solenoidal coils have demonstrated that the gap capacitance $C_G$ can be written as:[44,45]

$$C_G = \frac{\epsilon_0 \pi d_{coil}}{2\,\ln(d_{coil}/d_{REx})} \qquad (Eq.\,4)$$

The electric field outside the coil gives rise to a stray capacitance $C_0$ acting to increase the effective inductance of the coil. In addition, a resistance term $R_0(f)$ and a reactance term $X_0(f)$ account for the rest of the probe circuitry, eventually comprising tuning and matching components and circuit elements related to other NMR channels in the probe. The load impedance of the probe can then be written as:

$$Z(f) = R_0 + iX_0 + i\,2\pi f L_{eff} \qquad (Eq.\,5)$$

where the effective inductance of the coil is[45]:

$$L_{eff} = \frac{L_0}{1 - (2\pi f)^2 L_0 \left[C_0 + \dfrac{C_G(\varepsilon C_S + \varepsilon_R C_R)}{C_G + \varepsilon C_S + \varepsilon_R C_R}\right]} \qquad (Eq.\,6)$$

For a given set of parameters $L_0$, $C_0$, $C_G$, $C_S$ and $C_R$, the probe parameters $X_0(f)$ and $R_0(f)$ can be determined from the measured load impedance of the empty probe ($Z_E(f)$, as plotted in **Figure S2**), when no sample nor rotor is inserted in the coil:

$$Z(f;\,\varepsilon = \varepsilon_R = 1) = Z_E(f) \qquad (Eq.\,7)$$

The resonance frequency ($f_{res}$) of the probe can then implicitly be determined from the resonance condition, Im[$Z(f_{res})$] = 0.[46]

Non-general **Eq. 5** and **6**, valid for our modelled probe head, show how the impedance of a simplified probe head is altered by insertion of a dielectric material (the sample) into the coil, the impact of sample insertion on the resonance frequency being given by the resonance condition. Measuring the dielectric shift upon insertion of the sample is thus a way to measure the dielectric permittivity of the sample provided calibration curves converting between dielectric shift and sample permittivity can be constructed. Asides the dielectric permittivity of the sample, the dielectric shift will also depend on the stray capacitance, the gap capacitance, the rotor material, the settings of the tuning/matching network and on the settings of other probe channels, so that dielectric shift measurements and calibration curves should be done with identical full and empty rotors, taking the care not to adjust the tuning or matching networks between the measurements of the empty and full rotors.

From an experimental point of view, calibration curves can be constructed by inserting in the coil samples of known dielectric permittivity. **Figure 2** shows an example of such a curve, recorded on a 300 MHz Bruker probe head for samples with known dielectric permittivity. The curve was constructed by inserting different amounts of water in the NMR rotor (see Experimental section) and measuring the dielectric shift with respect to an empty rotor, while MAS spinning at 2 kHz. By spinning the rotor, a cylindrical water layer forms on the rotor walls. The increase in stray capacitance upon insertion of the water into the coil volume increases the effective inductance in the circuit, shifting the resonance frequency of the circuit to lower values. This results in a negative dielectric shift, increasing in magnitude with increasing sample permittivity.

By taking $C_S$ = 6.0 fF, $C_R$ = 5.1 fF and $C_G$ = 0.31 pF, as defined by **Eq. 2-4**, $L_0$ = 178 nH and $X_0(f)$ and $R_0(f)$ as defined by **Eq. 7**, the only free parameter defining the resonance frequency ($f_{res}$) of the probe head is the stray capacitance $C_0$. By fitting the modelled dielectric shift (calculated using the resonance condition 'Im[Z($f_{res}$)] = 0' for the empty and filled rotors) to the experimental data in **Figure 2**, a $C_0$ capacitance of 0.64 pF is found, with a R-squared value of 0.999. As a matter of comparison, for solenoidal coils of the same dimension of the one used in our modelled probe and when no conducting or dielectric materials are in the proximity of the coil, a stray capacitance $C_{stray}$ = 0.12 pF is expected related to the electric field lines that cross the region outside the coil.[47] The higher value of $C_0$ as compared to $C_{stray}$ can then be interpreted as a consequence of the presence of the stator and other conducting parts of the probe head around the receiver coil. In fact, the ratio $C_0/C_{stray}$ = 5.4 is consistent with the dielectric permittivity of boron nitride,[48] the base material of the stator employed in the probe head used in the reported measurements. The data is well described by the circuit model proposed in **Figure 2** (inset), showing that the interaction between sample and the field $E_1$ can successfully explain the observed dielectric shifts in NMR probe heads.

**Dielectric spectroscopy on alcohols**. The new methodology was validated by measuring the permittivity spectra of methanol, ethanol, 1-propanol, 1-pentanol and 1-octanol in the frequency range from 75 to 800 MHz (**Figure 3**). These aliphatic alcohols feature a strong dielectric relaxation process in this frequency range, with reported relaxation times of 0.0487, 0.158, 0.267, 0.652, 1.52 ns for increasing chain length.[49] The dielectric spectra acquired via NMR probe head calibration agrees with data acquired using standard dielectric spectroscopy techniques, thus showing the applicability of the method.

From **Figure 3**, it is possible to observe that a more scattered data was obtained for the permittivity spectra of methanol than for the other alcohols, a feature related to the decreased sensitivity of the method for increasing dielectric permittivity. In **Figure S3B** we show the sensitivity of the NMR probe head for permittivity measurements, defined by the slope of the calibration curves in **Figure S3A**. For low permittivity values, where the probe head is more sensitive, better accuracy for permittivity measurements is achieved and the spectra of 1-octanol, 1-pentanol, 1-propanol and ethanol closely follow previously reported data.[49,50] For methanol, the alcohol featuring the highest dielectric permittivity among the investigated ones, a more scattered permittivity spectrum was obtained, an effect related to the lower sensitivity of the probe head at higher dielectric permittivity.

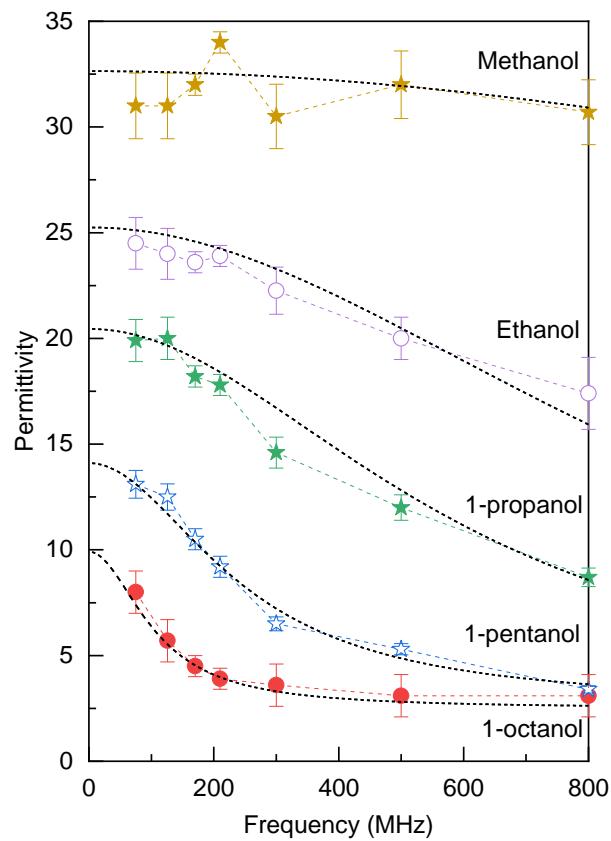

**Figure 3**. Dielectric permittivity spectra at 22 °C of aliphatic alcohols acquired using multi-diagnostic NMR/permittivity spectroscopy (symbols). Error bars are the standard deviation obtained from triplicate measurements. The black dashed lines represent reference literature data acquired by standard DRS techniques.[49,50]

**Dielectric permittivity of confined water.** As an example of the applicability of the developed methodology for investigating solvent-surface interactions and the physicochemical properties of confined systems, water confined in the micropores of an MFI-11.5 zeolite has been investigated. While defect-free all-silica MFI zeolites are highly hydrophobic materials, the framework defects and Brønsted acid sites (BAS) in Al-substituted zeolites such as MFI-11-5 facilitate water intrusion, with water being adsorbed in these hydrophilic sites.[14] The $^1$H NMR spectrum and the average permittivity of water in the micropores of MFI-11.5 is depicted in **Figure 4** for different hydration levels. In direct-excitation NMR spectra, adsorbed/confined and mobile species can be distinguished by their linewidths. Adsorbed/confined molecules are characterized by broad resonances, while mobile molecules are characterized by sharp resonances. This difference is induced by mainly two factors: the difference in rotational freedom that leads to a less efficient averaging of dipolar couplings that broaden the spectrum, and heterogeneity in adsorption sites. Up to *ca*. 100 % pore filling, the NMR spectra reveal mainly broad resonances corresponding to adsorbed water. After full pore filling is reached, sharp resonances corresponding to mobile water molecules start to appear on top of the broad resonance.

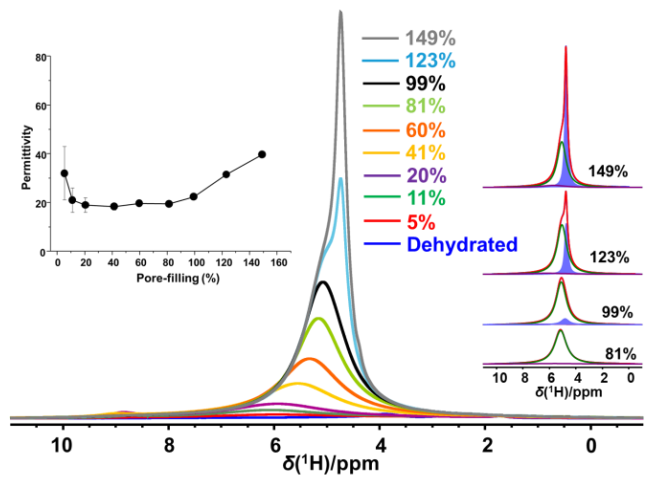

**Figure 4.** $^1$H MAS NMR spectra of MFI-11.5 zeolite exposed to different amounts of water. In the insets, (left) the average permittivity of water in the sample as a function of the degree of pore-filling and (right) spectral decomposition of $^1$H MAS NMR spectra at 81, 99, 123 and 149% pore-filling.

Interestingly, while only adsorbed/confined water is observed before full pore filling is reached, very low dielectric permittivity is observed for this adsorbed/confined water. As seen in the inset of **Figure 4**, at water loadings lower than *ca.* 80 % pore filling, average permittivity values as low as 19 are observed, reflecting the effect of nano-confinement on the dielectric permittivity of liquids[18] and the strong surface solvent interactions in the vicinity of the adsorption sites, which decreases the polarizability of the water molecules involved in the interaction[16]. As discussed by Parez *et al.*[16], surface proximity imposes restrictions in the geometric organization of water molecules thus lowering its the dielectric permittivity.[4,18] At the same time, the MFI-11.5 being a hydrophilic solid with variety of adsorption sites comprising BAS and defect sites (SiOH, AlOH) leads to strong adsorption of water, again restricting polarization.

Spectral decomposition allows quantification of the mobile water fraction revealed by the sharp resonance at 4.7 ppm (blue shaded area) highlighted in the inset of **Figure 4**. This water fraction features the same chemical shift observed for liquid water and appears in increasing quantities in the at hydration levels of 99, 123 and 149 % pore-filling as shown in **Table 1**. Assuming this mobile water fraction to feature a dielectric permittivity of 79, the same of bulk water, and the confined water to retain a dielectric permittivity of 19, the average dielectric permittivity of water in the sample can be calculated (see **Table 1**). The good agreement between experimental values and this calculation confirms the efficiency of the developed methodology in characterizing dielectric properties of confined water and other solvents. Since the degree of reorganization of the water network is dependent on the surface chemistry, determination of the change in dielectric properties of the intruded water offers a versatile methodology to characterize the influence of specific surface chemistry. This, in combination with the possibility of simultaneous access to quantitative atomic level information about speciation, interactions, dynamics, etc., from the NMR spectra[14,32,33], makes the developed methodology a versatile surface characterization technique for characterizing and generating structure-function relationships for wide variety of porous materials like zeolites, periodic mesoporous silicas, organo silicas, etc., with applications ranging from catalysis, adsorption and energy storage.

**Table 1.** Adsorbed and mobile water fractions determined via $^1$H NMR spectral decomposition and comparison of dielectric permittivity estimated from dielectric shift.

| Water loading (% pore-filling) | Fraction of water | | Dielectric constant | |
| --- | --- | --- | --- | --- |
| | Adsorbed | Mobile | Measured | Calculated* |
| 81.3 | 1 | 0 | | |
| 99.3 | 0.92 | 0.08 | 22.4 | 24.3 |
| 123 | 0.79 | 0.21 | 31.5 | 31.8 |
| 149 | 0.63 | 0.37 | 39.7 | 42.2 |

*Average assuming dielectric permittivity $\varepsilon_L$ = 79 for the liquid-like mobile fraction outside the pores and $\varepsilon_C$ = 19 for the confined water fraction.

## CONCLUSIONS

While the magnetic interactions between coil and sample in an NMR probe head allow characterization of atomic level interactions thought the acquisition of NMR spectra, the electric interactions in the system allow to simultaneously probe the dielectric properties of the sample. Via the calibration of the NMR probe head, dielectric permittivity measurements on liquid and solid systems can be performed in the NMR frequency range. Comparison between dielectric permittivity data acquired via this calibration and via other traditional EIS/DRS techniques demonstrates the accuracy of the developed methodology. Multi-diagnostic NMR/dielectric permittivity characterization of water confined in the micropores of an MFI-type hydrophilic zeolite with Si/Al ration of 11.5 revealed the drastic decrease in the dielectric permittivity of water due to confinement and the interaction with the pore wall. Multi-diagnostic *in situ* NMR and permittivity spectroscopy is expected to benefit physicochemical characterization not only of confined water systems, but also of electrolyte and sensor materials.

## ASSOCIATED CONTENT

**Supporting Information**. Nomenclature adopted for the input and load impedances. Electric delay determination. Dielectric permittivity calibration curves. Discussion on the model circuit. This material is available free of charge via the Internet at http://pubs.acs.org.

## AUTHOR INFORMATION

Corresponding Author

* Eric Breynaert - Centre for Surface Chemistry and Catalysis – Characterization and Application Team (COK-KAT) & NMRCoRe, KU Leuven, Celestijnenlaan 200F Box 2461, 3001-Heverlee; orcid.org/0000-0003-3499-0455; Email: Eric.Breynaert@kuleuven.be.

**Funding Sources**

This work has been funded by the European Research Council through an Advanced Research Grant under the European Union's Horizon 2020 research and innovation program (No. 834134 WATUSO). A.M. thanks H2020 Marie Skłodowska-Curie Actions for a postdoctoral fellowship (H2E; 101063656). E.B. acknowledges joint funding by the Flemish Science Foundation (FWO;

G083318N) and the Austrian Science Fund (FWF) (funder ID 10.13039/501100002428, project ZeoDirect I 3680-N34). NMRCoRe is supported by the Flemish government as International Research Infrastructure (I001321N: Nuclear Magnetic Resonance Spectroscopy Platform for Molecular Water Research) and received infrastructure funding from the Flemish government, department EWI via the Hermes Fund (AH.2016.134).

**Notes**
There are no conflicts to declare.

**ACKNOWLEDGMENTS**